\documentclass[prb,preprint,aps]{revtex4} 
\usepackage{epsfig} 
\usepackage{amsmath} 
\usepackage{amssymb} 
\usepackage{graphicx} 
\usepackage{multirow} 
\usepackage[usenames]{color} 
\usepackage[english]{babel} 

\newcommand{\be}{\begin{equation}}
\newcommand{\ee}{\end{equation}}
\newcommand{\bea}{\begin{eqnarray}}
\newcommand{\eea}{\end{eqnarray}}

\begin{document} 
\title{Magnetization, Maxwell's Relations and the Local Physics of
  Th$_{1-x}$U$_x$Ru$_2$Si$_2$         } 
\author{Anna T\'oth$^1$, Premala Chandra$^1$, Piers Coleman$^1$,  Gabriel Kotliar$^1$ and Hiroshi Amitsuka$^2$} 
\affiliation{$^1$ Center for Materials Theory, Department of  
Physics and Astronomy, Rutgers University,  
Piscataway, NJ 08854} 
\affiliation{$^2$ Department of Physics, Hokkaido University, Sapporo 060-0810, 
Japan} 
\date{\today} 
\begin{abstract} 
The dilute Kondo compound, 
Th$_{1-x}$U$_x$Ru$_2$Si$_2$,         
displays non-Fermi
liquid behavior but no zero-point entropy; it thus appears to elude 
description by known single-ion models.  It may also provide a clue to 
the underlying local degrees of freedom in its dense counterpart,
URu$_2$Si$_2$.  Here we use high-resolution magnetization studies to 
cross-check the thermodynamic consistency of previous experiments.
Measurement of the field-dependence of the temperature-scale, $T_F(H)$,
associated with Fermi liquid behavior probes the nature of the underlying
impurity fixed point.  We find that $T_F(H)$ grows linearly with applied field,
in contrast to the quadratic form expected for the two-channel Kondo model.
We use a scaling argument to show that the observed behavior of $T_F(H)$
is consistent with the absence of zero-point entropy,
suggesting novel impurity behavior in this material. More generally, 
we suggest the field-magnetization as a probe of single-ion 
physics and make predictions for its behavior in other actinide compounds. 
\end{abstract} 
\maketitle 
The non-Fermi liquid (NFL) physics of the dilute Kondo  
compound Th$_{1-x}$U$_x$Ru$_2$Si$_2$ (TURS) is an outstanding problem in heavy fermion 
materials. The local degrees of freedom responsible for NFL behavior in TURS 
are widely believed to provide the Hilbert space for   
the hidden order in its dense counterpart, 
URu$_2$Si$_2$ (URS); thus understanding of the dense and the dilute systems 
may be closely linked.  Furthermore, although TURS has been extensively  
studied experimentally,\cite{Amitsuka94,Amitsuka00} its unusual physics has eluded description by an 
established single-ion model known to display NFL behavior. 
For example, a well-studied mechanism for NFL in an impurity system is 
provided by the two-channel Kondo model (2CKM), where competition between 
the channels results in quantum critical behavior accompanied by 
a fractional zero-point entropy (FZPE).\cite{Cox96} 
It has been proposed that this physics is realized in a number of  
heavy fermion impurity systems 
characterized by quadrupolar or non-Kramers doublet ground-states;\cite{Cox87}
TURS was initially thought to be an excellent candidate.   
In particular experiments\cite{Amitsuka94,Amitsuka00} indicate 
$\gamma (\equiv \frac{c_P}{T}), \chi \sim \ln T$ at low temperatures, 
and application of a magnetic field ($H$) drives the system into a Fermi 
liquid with $\gamma, \chi \sim \log H$.  However the 2CKM proposal fails 
in a crucial way:  since application of $H$ quenches the FZPE, the 2CKM 
predicts a field-induced Schottky anomaly in the specific heat ($c_P$). 
However, in contrast to the situation\cite{Seaman91} in Y$_{1-x}$U$_x$Pd$_3$, 
this is not observed\cite{Amitsuka94,Amitsuka00} in TURS and 
so the FZPE predicted by the 2CKM\cite{Shimizu99} appears to be absent.  
Here we return to this problem
in TURS spurred by renewed interest in the dense system URS.
We show that high-resolution magnetization studies provide a cross-check
on the thermodynamic consistency of previous specific heat experiments.  More specifically,  
measurements of the field-dependence of the temperature-scale, $T_F(H)$, associated with
Fermi liquid behavior probes the nature of the underlying impurity fixed point.
We therefore study whether the failure to observed the FZPE in TURS is an experimental
issue or whether it indicates the presence of a fundamentally new class of impurity
behavior.                              
 
The magnetization, in conjunction with Maxwell's thermodynamic relations, 
can be used to cross-check specific heat measurements.
For a system with magnetic moment $\,m\,$, Maxwell's relation 
\be
\frac{\partial^2 F}{\partial H\partial T}\, =\, 
-\,\frac{\partial m}{\partial T}\, =\, -\,\frac{\partial S}{\partial H} 
\ee
leads to 
\be
\Delta S\, =\,\int \int\,\frac{\partial \chi}{\partial T}\, dH' dH 
\,=\,\int\, \frac{\partial m}{\partial T}\, dH,
\ee
so that
\bea
S(T,H)\, -\, S(T,0)& = &\,\int_0^H  \,\frac{\partial m(T,H')}
{\partial T }\, dH' 
\eea
and 
\bea\label{Max1}   
-\,S_{ZP}\, \equiv\,-\,\lim_{T\to0}S(T,0)\,=\,  \lim_{T \rightarrow 0}\,\int_0^H \, \frac{\partial m(T,H')}{\partial T}\, dH', 
\eea
resulting in 
\bea\label{Max2} 
\lim_{T \rightarrow 0}\,\frac{\partial m}{\partial T} \left[sgn(H)\right]& \sim & - \,2 S_{ZP}\, \delta (H)\,, 
\eea
where $sgn(H)$ is included in (5) to ensure that both sides of this
equation are even in field.
Since high-resolution magnetization measurements do not have the subtraction issues
associated with $c_P$ experiments, (\ref{Max2}) can be used to cross-check the ZPE result for TURS.

\begin{figure}[ht]
  \includegraphics[width=0.7\columnwidth,clip]{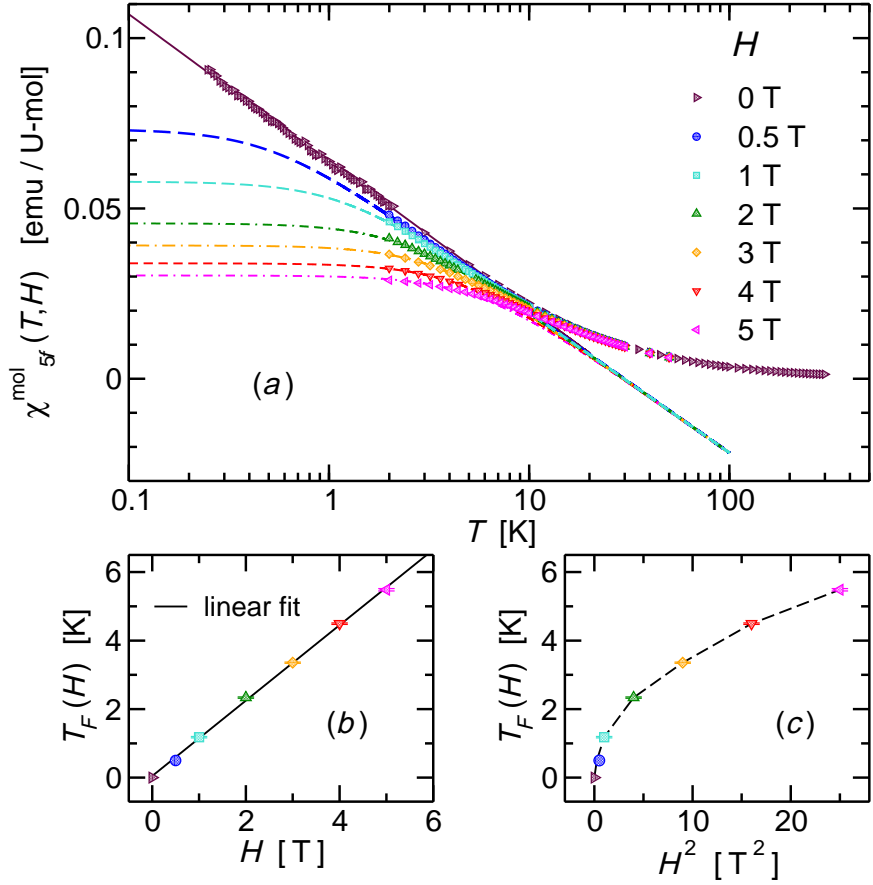}
  \caption{(color online) $(a)$ Molar susceptibility, $\chi_{5f}^{mol}$ of U in Th$_{1-x}$U$_x$Ru$_2$Si$_2$
    for $\,x=0.03$ as the function of temperature
    for applied fields ranging from $\,H\,=\,0$ to 5 T. Lines
    indicate the best fit to the data below 6 K with the form $\chi_0\, \ln\left[{T_K}\,/\,{\sqrt{T^2 \,+\,
    T_F(H)^2}}\right]\,$ with
    $\,\chi_0\,=\,0.018\,\pm\,10^{-3}\,{\textrm{emu}}\,/\,{\textrm{U-mol}}$
    and $\,T_K\,=\,29.8\,\pm\,0.3\,$K. The crossover
    scale, $T_F(H)$ is plotted on linear $(b)$ and quadratic $(c)$ scales. The
    uncertainty  of the fit for the parameter $\,T_F(\,H\,=\,0\,\textrm{T})\,=\,0$
    is much smaller than the symbol size, while the
    value, $\,T_F(\,H\,=\,0.5\,\textrm{T})\,=\,0.5\,${K} has been fixed. 
  }
  \label{fig:molar_suscept}
\end{figure}

In Figure \ref{fig:molar_suscept}$(a)$ we show the $\,T$-dependence of
$\, \chi_{5f}^{mol}\,$, the molar susceptibility of U in
Th$_{1-x}$U$_x$Ru$_2$Si$_2$ for $\,x\,=\,0.03\,$ in magnetic fields
between $ H= 0 $ and 5 T.  
Noting the
logarithmic behavior in $\chi$ close to the quantum critical point at
zero temperature and magnetic field, 
we can model it with the simple expression 
\bea 
\chi_{5f}^{mol}(T,H) =
-\,\chi_0\, \ln\frac{\sqrt{T^2 + T_F(H)^2}}{T_K}.
\eea 
Fitting this form to the data below
$\,T\,=\,6\,$K, we obtain
$\,\chi_0\,=\,0.018\,\pm\,10^{-3}\,\frac{\textrm{emu}}{\textrm{U-mol}},\,T_K\,=\,29.8\,\pm\,0.3\,$K.
Figures \ref{fig:molar_suscept}$(b)$ and $(c)$ show the
field-dependent crossover temperature, $T_F(H)$ on linear and
quadratic scales. Its markedly linear magnetic field-dependence
contrasts strikingly with the quadratic behavior characterizing the
2CKM. The crossover scale $k_{B}T_{F} (H)$ is quantitatively the
magnitude
of a Zeeman energy 
\begin{equation}\label{}
k_{B}T_{F} (H) = (gS)\mu_{B}H \equiv k_B h,
\end{equation}
with a g-factor of $gS= 1.6$ where we have introduced the reduced
field, $h \equiv (gS)\mu_B H/ k_{B}$. 
The appearance of a Zeeman splitting in the temperature dependence of
the magnetic susceptibility is an indication that the underlying
single-ion ground-state of TURS is a magnetic multiplet. The value $gS=1.6$ would correspond to a Ising magnetic
moment $gS \times \mu_{B}= 1.6 \times\mu_{B}$ a value consistent with
previous estimates for a magnetic $\Gamma_{5}$ doublet\cite{Amitsuka94}.
In the hexadecapolar Kondo effect scenario\cite{Haule10}, 
$g = 3.2 \cos(\phi)$, which then sets the mixing angle to be $\phi = \frac{2 \pi}{3}$ in the dilute limit.


Since the phenomenological fit (6) to the susceptibility data can be
written as 
\bea
\chi &\propto& -\, \ln\left[T^2\, +\, T_F(H)^2\right] \,+\,cst\,,\\
\ln\left[T^2\, +\, T_F(H)^2\right]&\propto& {\rm Re}\, \ln \left[T_F(H)\, -\, i\,T\right] 
\eea
this leads to
\bea
\frac{\partial \chi}{\partial T} &\propto &{\rm Im}\, \frac{1}{T_F(H) \,-\, i\,T}.
\eea
Replacing $T_F(H)$ with $h$ and using the Maxwell relation (2),  we obtain
a phenomenological form for the entropy
\bea
\Delta S&\propto& {\rm Im}\left[(h\, -\, i\,T) \ln\left({h\, -\, i\,T}\right)\, +\, i\,T \ln
  \left({-\,i\,T}\right)\,-\,h\log(-\,i)\right]\,
\label{entropy}
\eea
which is a regular smooth function for the full phase region that includes
$h > T$ and $T > h$.  More explicitly
\bea 
\lim_{H\rightarrow 0}\, \lim_{T\rightarrow 0} \,\Delta S = \lim_{T\rightarrow 0}\, \lim_{H\rightarrow 0}\, \Delta S =0, 
\eea
indicating that there is no order-of-limits issue, no irregularity
and thus no zero-point entropy; this is consistent with 
the previous specific heat results.\cite{Amitsuka94,Amitsuka00}

We can also understand this absence of zero-point entropy using a more general 
scaling argument where we assume a regular scaling function.  
From experiment, we obtain 
\bea
\chi& \sim& h^{-\nu}, 
\eea
where for TURS $\nu = 0$.  We can write a general 
scaling form for the free-energy 
\bea
F &\sim& h^{2-\nu} \Phi\left(\frac{T}{h^\eta}\right), 
\eea
where $\eta$ refers to the field-dependence of the crossover scale, $T_F(H)$, and 
we assume that $\Phi(x)$ is a slowly varying function of $x$. 
Then 
\bea
F &\sim & T^{\frac{2-\nu}{\eta}}\, \Phi_1 \left(\frac{h}{T^{\frac{1}{\eta}}}\right), 
\eea
and 
\bea
S\, =\, -\,\frac{\partial F}{\partial T} \,=\, T^{\frac{2-\nu}{\eta} - 1}\, \Phi_2 \left(\frac{h}{T^{\frac{1}{\eta}}}\right) \,.
\eea
In order to have a finite zero-point entropy, the exponent of the temperature must be zero; then we must have 
\bea\label{eq:scaling_relation}
\eta& =& 2 - \nu \,.
\eea
We see that to have ZPE for $\nu = 0$, $\eta = 2$ is required; thus the absence of an observed zero-point entropy is consistent 
with the measured $\eta = 1$. The scaling relation \eqref{eq:scaling_relation}
is realized in the multichannel Kondo models.
There the exponents, $\,\eta\,$
and $\,\nu\,$ are related to the number of screening channels, $\,k\,$  by
\cite{Andrei84,Wiegmann85,Affleck91}
\bea
\eta\,=\,1\,+\,\frac 2 k\,,\quad\quad\quad
\nu\,=\,1\,-\,\frac 2 k\,.
\eea

\begin{figure}[ht]
  \includegraphics[width=0.7\columnwidth,clip]{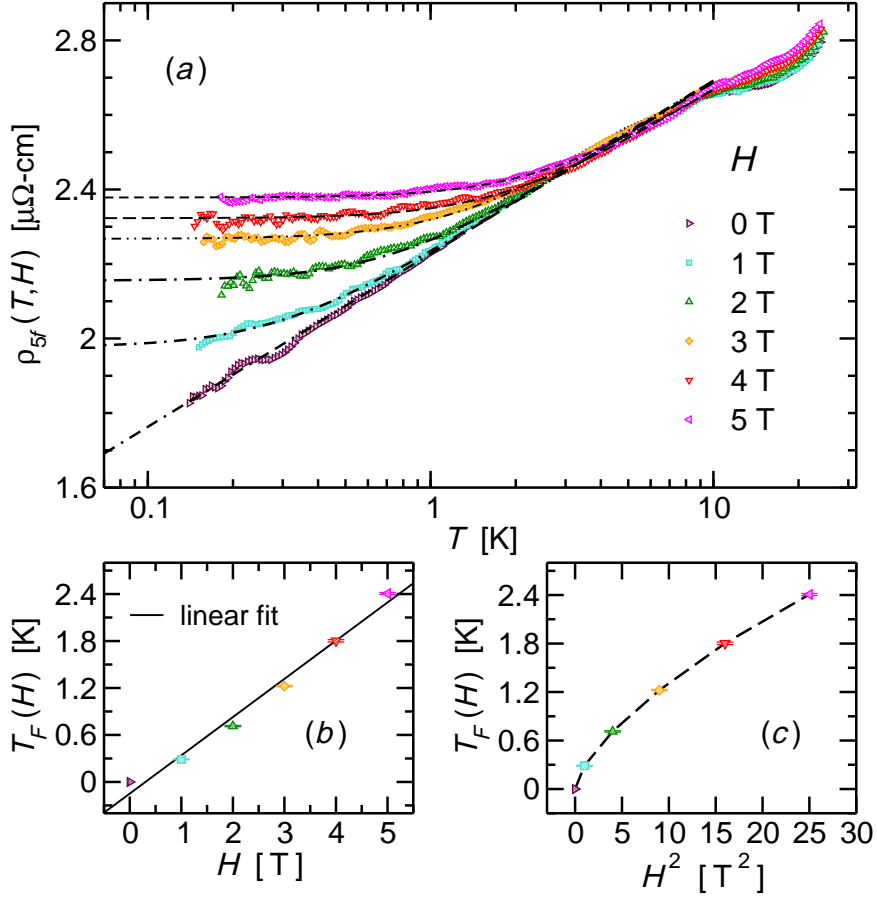}
  \caption{(color online) $(a)$ Electrical
    resistivity for currents along the $a$-axis  in
    Th$_{1-x}$U$_x$Ru$_2$Si$_2$ for $\,x\,=\,0.03\,$ ($\,\rho_{5f}(T,H)\,$) as the function
    of $\,T\,$ for various fields between 0 and 5
    T. Dashed lines indicate the best fits to the data below 9.5 K with the
    form  
    $\,\rho_0\,\ln\left({\sqrt{T^2\,+\,T_R(H)^2}}\,/\,{T_0}\right)\,$ with
    $\,\rho_0\,=\,0.201\,\pm\,.002$ $\mu\Omega$-cm and  $\,T_0\,
    =\,(1.6\,\pm\,0.1)\,\times\, 10^{-5}\,$ K.  The crossover scale,
    $\,T_F(H)\,$ vs  $\,H\,$ and $\,T_F(H)\,$ vs $\,H^2\,$ is shown 
    on plots $(b)$ and $(c)$. There the standard errors coming from the
    non-linear fit are
    smaller than the symbol size. 
  }
  \label{fig:resist}
\end{figure}

Further support for the $\frac{H}{T}$ scaling comes from resistivity data
shown in Figure {\ref{fig:resist} again for a TURS sample with $x = 0.03$.  We have fit
it with the form
\bea
\rho = \rho_0 \ln\left({\sqrt{T^2\,+\,T_R(H)^2}}\,/\,{T_0}\right)\,
\eea
where $T_R(H)$ is the dynamical crossover scale for Fermi liquid behavior
to develop.  Once again we see a linear $H$-dependence of this crossover
scale; because we only have at most two decades of data we do not present
it as a scaling plot.  The fact that $T_R(H)$ and  $T_F(H)$ are proportional
to each other but are not equal reflects 
that the $H/T$ scaling functions 
associated with thermodynamics and transport are most likely different.  More field-dependent measurements of thermodynamic
responses and resistivity at low fields and temperatures on TURS would
provide more specifics about the nature of these scaling functions.

\begin{figure}[ht]
  \includegraphics[width=0.85\columnwidth,clip]{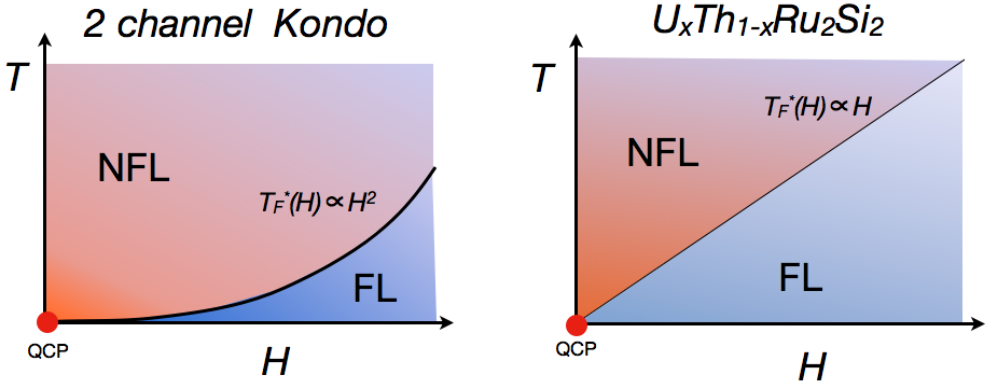}
  \caption{(color online) Schematic phase diagram for (a) two channel
  Kondo model compared with experimentally determined phase diagram  for
 (b) U$_{x}$Th$_{1-x}$Ru$_{2}$Si$_{2}$.
  }
  \label{fig:schematic}
\end{figure}

It is intriguing that the zero-field properties of TURS significantly
overlap with those of the 2CKM, but that application of a magnetic field yields
two very different field-induced Fermi liquids: in the TCKM the
Fermi temperature grows quadratically with field whereas in TURS it
is linear in field, as illustrated in  Figure (\ref{fig:schematic}).
The 2CKM has a residual zero-point entropy, yet TURS has none;
moreover these two features can be related by scaling arguments.
Taken together these clear differences suggest a new kind of impurity
fixed point behavior with a novel kind of non-Fermi liquid behavior.
What kind of Kondo model can account for this new physics?

Kondo models involve
small localized Hilbert spaces describing spin and orbital degrees of
freedom  coupled to one or more conduction baths.
The generic ground-state of an asymmetric Kondo model is a Fermi
liquid. However,  if the competing screening channels are symmetry-equivalent, 
then non-Fermi liquid behavior and a
residual entropy result.  Is there a  deviation from perfect channel symmetry
that is at once strong enough to destroy the zero-point entropy whilst
remaining weak enough to preserve some type of non-Fermi liquid behavior?  
In two channel Kondo models, 
deviation from channel symmetry on the Fermi surface 
immediately leads to Fermi liquid behavior. In principle this leaves open
the possibility of a {\sl marginal} channel
asymmetry that is absent at the Fermi surface but grows as one moves
away from it. For example, in the $\Gamma_5$
scenario,\cite{Amitsuka94,Ohkawa99} one isospin direction is odd under
time-reversal whereas the other two are even.  Thus there is weaker
symmetry protection than in the usual 2CKM scenario,\cite{Haule09} and
further investigation is necessary to see whether marginal channel
asymmetries exist here.  We also note that an intermediate asymptotic
regime with $T_F(H) \propto H$ can be obtained within the
hexadecapolar Kondo scenario provided that the crystal-field splitting
between the $\Gamma_1$ and the $\Gamma_2$ singlets is small and an
intermediate-coupling condition is obeyed.\cite{Toth10}
 
In conclusion, we have used high-resolution magnetization measurements
to confirm the absence of a zero-point entropy 
in TURS. Exploiting the fact that an
applied field restores Fermi liquid behavior in TURS, we find that the
field-dependent Fermi temperature $T_F(H)$ scales {\sl linearly} with
field rather than the quadratic behavior expected for the 2CKM.  Since
this technique does not depend on subtraction issues, it would be
interesting to apply it to various impurity systems previously found to
display quadrupolar Kondo behavior\cite{Seaman91} where we expect $T_F
(H) \sim H^4$ or $T_F(s) \sim s^2$ where $s$ is strain. Of particular
interest is the quadrupolar Kondo
candidate\cite{Kawae06} $Pr_xLa_{1-x}Pb_3$ for $x \le 0.05$ where 
no ZPE has been observed.  Finally, we would like to encourage more
low-field and low-temperature measurements on TURS to learn more about
the nature of its underlying impurity fixed point.  

We acknowledge helpful discussions with N. Andrei, C. Bolech, D.L. Cox
and M. B. Maple.  This work was supported in part by NSF DMR-0906943
(G. Kotliar and A. T\'oth), NSF NIRT-ECS-0608842 (P. Chandra) and DOE
DE-FG02-99ER45790 (P. Coleman).

\end{document}